\documentclass[twocolumn]{aa}

\usepackage{graphicx}
\usepackage{natbib}
\usepackage{amssymb}

\bibpunct{(}{)}{;}{a}{}{,}
\newcommand{\be}{\begin{equation}}
\newcommand{\ee}{\end{equation}}
\begin{document}

\title{Optical microvariability of EGRET blazars\thanks{Based on
observations made at the Complejo Astron\'omico El Leoncito, which is
operated under agreement between CONICET and the National Universities
of La Plata, C\'ordoba, and San Juan.}}

\author{Gustavo E. Romero \inst{1,}\thanks{Member of CONICET},
 Sergio A. Cellone \inst{2},
 Jorge A. Combi \inst{1, }$^{\star\star}$,
Ileana Andruchow \inst{1}}

\offprints{Gustavo E. Romero}

\institute{Instituto Argentino de Radioastronom\'{\i}a, C.C.5,
(1894) Villa Elisa, Buenos Aires, Argentina \and
Facultad de Ciencias Astron\'omicas y Geof\'{\i}sicas, UNLP, Paseo del
Bosque, 1900 La Plata, Buenos Aires, Argentina}

\date{Received 15 January 2002 / Accepted 15 May 2002}

\titlerunning{Microvariability of EGRET blazars}
\authorrunning{G.E. Romero et al.}

\abstract{We present results of a photometric CCD study of the
incidence of microvariability in the optical emission of a sample
of 20 blazars detected at gamma-ray energies by the EGRET
instrument of the Compton Gamma-Ray Observatory. We have observed
strong outbursts in some sources, but many others displayed no
significant variability on timescales of hours. The typical
minimum timescale results to be of $\sim$ several hours, not tens
of minutes as claimed by some authors. The duty cycle for optical
intranight microvariations of gamma-ray blazars, as estimated from
our observations, seems to be $\sim 50$ \%, lower than what is
usually assumed. For night-to-night variations, instead, the duty
cycle approaches to what is observed in radio-selected BL Lacs and
flat-spectrum radio quasars (i.e. $\sim70$ \%).
}

\maketitle

\section{Introduction}

The rapid changes in the optical brightness of blazars, typically with
timescales of less than a single night, are well known. The existence
of this phenomenon, usually called {\em microvariability}, was only
accepted by the astronomical community after the advent of modern CCD
photometry in the 1980s \citep{MCG89}, despite the existence of
previous reports \citep[e.g.,][]{R70}. In its most extreme
manifestations the microvariations of blazars can reach values of more
than 100\% in less than 24 hours \citep[e.g.,][]{RCC00a}. The
origin of such an amazing behaviour is yet not clear.

The duty cycles (i.e., the fraction of time spent by a given
source or group of sources displaying microfluctuations) for
different classes of AGNs are not well known, but is usually
thought that in radio-loud quasars (RLQs) and radio-selected BL
Lac objects (RBLs) they are higher than those presented by X-ray
selected BL Lacs (XBLs) and radio-quiet quasars (RQQs)
\citep[e.g.,][]{HW96,HW98,RCC99,GK00,QTF00,QTF02}. Duty cycles of RBLs and
RLQs are estimated to be $\sim 70$\%, whereas the corresponding
values for XBLs and RQQs seem to be $\sim 30$\% and $\sim 7$\%,
respectively \citep{RCC99}. The high optical duty cycles displayed
by strong flat-spectrum radio sources seem to be a consequence of
the presence of relativistic jets oriented close to the line of
sight in these objects. The microfluctuations could arise from
interactions of relativistic shocks with small features in the
parsec-scale jets \citep[e.g.,][]{Q91,Q00,R95,K99}. The lower duty
cycles of XBLs could be a consequence of the stronger magnetic
fields in these objects \citep[e.g.,][]{RCC99}. In RQQs, the
microvariations are possibly related to instabilities and orbiting
hot spots in the accretion disks \citep[e.g.,][]{MW93}, and the
duty cycles are perhaps reflecting the incidence of these
phenomena in the innermost part of the disks.

If the scenario outlined above is basically correct, one could expect
that gamma-ray blazars, the most energetic subclass of RL objects,
should present the highest duty cycles of all AGNs. In fact, some
recent monitoring campaigns by \citet{X99,X01,X02} seem to suggest high
duty cycles in a sample of northern blazars that have been detected by
the Energetic Gamma Ray Experiment Telescope (EGRET) of the late
Compton Gamma-Ray Observatory.  However, comparison with duty cycles
presented by other classes of objects requires uniform procedures for
data analysis and error control.  Very recently, \citet{CRC00} have
demonstrated through a combination of observations and photometric
simulations that small seeing fluctuations can be an important source
of spurious microfluctuations in differential photometry due to
variable contamination by light from host galaxies. Comparison of
results obtained by differential photometry with different instruments
and different photometric apertures should be treated with extreme
care. Some recent contradictory claims in microvariability research
could be due to an incorrect treatment of the errors in this kind of
observations \citep[see][for a detailed discussion]{CRC00}.

In this paper we present results of an extensive study of the
incidence of microvariability in a sample formed mostly by southern
gamma-ray blazars.  Observational technique and error control follow
the guidelines given by \citet{CRC00} and, consequently, our results
can be compared with those obtained by \citet{RCC99} for other types
of AGNs, since both studies were conducted with the same instrument
and identical procedures for data analysis.

In the next section we shall present our sample and describe the
observations and the data analysis. We then present our main results
in Sect.~\ref{s_mr}. The duty cycle for EGRET blazars is estimated in
Sect.~\ref{s_dc} and compared with other results found in the
literature. We close in Sect.~\ref{s_disc} with a brief discussion on the
origin of microvariability in blazars.

\section{Observations and data analysis \label{s_obs}}

The Third EGRET Catalog \citep{H99} lists 271 point-like gamma-ray
sources.  Of these sources 66 have been positively identified with
blazars, which are usually strong flat-spectrum radio sources
\citep[e.g.,][]{M97}. We have selected a sample of 20 of these blazars
that satisfy the following criteria: 1) they are located at
declinations lower than $+20^{\circ}$, and 2) they are brighter than
magnitude $m_V=19.0$. All these sources fall within the categories of
RBLs, XBL or RLQs, where we include both highly polarized QSOs
(HPQs) and optically variable violent QSOs (OVVQs). The sample is
presented in Table~\ref{t_obj}, where we list, from left to right, the
name of the object, the coordinates (RA and DEC) at J2000.0, the
redshift, the magnitude in the $V$ band,
the object type, the name of the corresponding source in the Third
EGRET Catalog, the averaged gamma-ray flux in units of
$10^{-8}$\,ph\,cm$^{-2}$\,s$^{-1}$, its error, and the high-energy
photon spectral index $\Gamma$.

\begin{table*}[tb]
\caption{Sample. The EGRET averaged (P1234) flux $F$ is
in units of 10$^{-8}$\,ph\,cm$^{-2}$\,s$^{-1}$.}

\begin{tabular}{lcrccclrcc}

\noalign{\medskip}
\hline
\noalign{\smallskip}

~~Object & $\alpha_{2000.0}$ & $\delta_{2000.0}$ & $z$ &
 $m_\mathrm{V}$ & Type & EGRET Name & $F$~~ & $\Delta F$ &
 $\Gamma\pm\Delta\Gamma $ \\

  & hs min s & $^{\circ} \quad ' \quad '' $& & & & ~~~~~3EG & & & \\

\noalign{\smallskip}
\hline
\noalign{\smallskip}

0208$-$512 & 02 10 46.2 & $-$51 01 02 & 1.003 &16.9  &RLQ &
J0210$-$5055 &85.5 &4.5 & $1.99\pm0.05$ \\

0235$+$164 &  02 38 38.9 & $+$16 36 59  & 0.904 & 19.0&RBL &
J0237$+$1635 & 25.9 & 3.7 & $1.85\pm0.12$ \\

0521$-$365 & 05 22 58.0 & $-$36 27 31 & 0.055 & 14.5 &RBL &
J0536$-$3626& 15.8& 3.5 & $2.63\pm0.42$ \\

0537$-$441 &05 38 50.4 & $-$44 05 09&0.894 & 15.5 &RBL &
J0540$-$4402 & 25.3 & 3.1 & $2.41\pm0.12$ \\

1226$+$023 & 12 29 06.7 & $+$02 03 09 & 0.158 & 12.8 &RLQ &
J1229$+$0210&15.4 & 1.8 & $2.58\pm0.09$ \\

1229$-$021& 12 32 00.0 & $-$02 24 05 & 1.045 & 17.7 &RLQ &
J1230$-$0247 &6.9 &1.5 & $2.85\pm0.30$ \\

1243$-$072 & 12 46 04.2 & $-$07 30 47 & 1.286 & 19.0 &RLQ &
J1246$-$0651& 9.8 & 2.1 & $2.73\pm0.17$ \\

1253$-$055 & 12 56 11.2 & $-$05 47 22 & 0.538 &17.8  &RLQ &
J1255$-$0549 &74.2 &2.8 & $1.96\pm0.04$ \\

1331$+$170 & 13 33 35.8 & $+$16 49 04 & 2.084 & 16.7 &RLQ &
J1329$+$1708 &4.4 &1.6 & $2.41\pm0.47$ \\

1334$-$127 & 13 37 39.8 & $-$12 57 25 & 0.539 &17.2 &RLQ &
J1339$-$1419 &5.5 &1.9 & $2.62\pm0.42$ \\

1424$-$418 & 14 27 56.3 & $-$42 06 19 & 1.522 & 17.7 &RLQ &
 J1429$-$4217&11.9&2.7 & $2.13\pm0.21$ \\

1510$-$089 & 15 12 50.3 & $-$09 06 00 & 0.361 & 16.5 & RLQ &
J1512$-$0849 &18.0 &3.8 & $2.47\pm0.21$ \\

1606$+$106 & 16 08 46.2 & $+$10 29 08 & 1.226 & 18.5 &RLQ &
J1608$+$1055 & 25.0 &4.5 & $2.63\pm0.24$ \\

1622$-$297 & 16 26 06.0 & $-$29 51 27 & 0.815 &20.5 &RLQ &
J1625$-$2955 &47.4 &3.7 & $2.07\pm0.07$ \\

1741$-$038 & 17 43 59.0 & $-$03 50 05 & 1.054 & 18.6 &RLQ &
J1744$-$0310 & 11.7 & 3.3 & $2.42\pm0.42$ \\

1933$-$400 & 19 37 16.2 & $-$39 58 02 & 0.966 & 18.0 &RLQ  &
J1935$-$4022 & 8.5 & 2.7 & $2.86\pm0.40$ \\

2022$-$077 & 20 25 40.6 & $-$07 35 52 & 1.388 & 18.5 &RLQ &
J2023$-$0836& 21.2 & 3.5 & $2.38\pm0.1$7 \\

2155$-$304 & 21 58 52.1 & $-$30 13 32 & 0.116 & 13.1 & XBL &
J2158$-$3023 & 13.2 & 3.2 & $2.35\pm0.26$ \\

2230$+$114 & 22 32 36.4 & $+$11 43 51 & 1.037 & 17.3 &RLQ &
J2233$+$1140& 19.2 & 2.8 & $2.45\pm0.14$ \\

2320$-$035 & 23 23 32.0 & $-$03 17 05 & 1.411 & 18.6 &RLQ &
J2321$-$0328&$<6.0$ & \ldots & \ldots \\

\noalign{\smallskip}
\hline

\end{tabular}
\label{t_obj}
\end{table*}

Objects of this sample were monitored repeatedly during several
observing sessions with the 2.15-m CASLEO telescope at El Leoncito,
San Juan, Argentina, from July 1997 till July 2001. The instrument was
equipped with a liquid-nitrogen-cooled CCD camera, using a Tek-1024
chip with a read-out-noise of 9.6 electrons and a gain of 1.98
electrons adu$^{-1}$. This is the same camera used by \citet{RCC99}
for a comparative study of duty cycles of radio-loud and radio-quiet
QSOs. The unvignetted field was approximately 9 arcminutes in
diameter, and consequently large enough as to contain several stars
for comparison and variability control. Exposures varied from $\sim 1$
minute to $\sim 5$ minutes, according to the brightness of the object
and the observing conditions. The quality of each night is coded with
a parameter $q$, according to the following scheme: photometric
($q=1$), clear but seeing not very good ($q=2$), thin cirrus ($q=3$),
thick cirrus or partially cloudy ($q=4$). This information is
given in column 3 of Table~\ref{t_results}. It can be seen that $2/3$
of our nightly lightcurves were obtained under good to very good
atmospheric conditions.

The observations were made with a Johnson $V$ filter. The CCD frames
were bias-subtracted and flat-fielded using median-averaged dome
flats, which resulted in a flat-field accuracy typically better than
$\sim 0.5 - 1.0 \%$ of the sky level. Standard stars from Landolt's
\citeyearpar{L92} fields were also observed each night for calibration
to the standard system.

The data reduction was made with IRAF\footnote{IRAF is distributed by
the National Optical Astronomy Observatories, which are operated by
the Association of Universities for Research in Astronomy, Inc., under
cooperative agreement with the National Science Foundation.}  software
package running on a Linux workstation.  Differential photometry was
then made with the aperture routine \textsc{apphot}.  Each set of data for each
object was always reduced with the same aperture radius, which was
determined taking into account the apparent size and brightness of the
host galaxy, in accordance with the recommendations by
\citet{CRC00}. The presence of neighbour stars was also taken into
account, and in a few cases they had to be subtracted using
\textsc{daophot} before performing aperture photometry. In any
case, the aperture diameter was never lower than $\sim 2.5$ times the
seeing FWHM.

Cross-checked non-variable stars in the field (with apparent
magnitudes as close as possible to that of the target object) were
divided in two groups, averaged, and used for comparison and control
as in \citet{RCC99}. Differential lightcurves were then computed as
target minus averaged comparison. In addition to the pure photometric
errors, spurious variability was pondered through the scatter of the
comparison minus control stars. The errors during a typical
microvariability session (1 night) were mostly in the range $\sim
0.002-0.008$ mag, although in some particular cases they reached
values of $\sim\,0.01$ mag. Coordinates and magnitudes for the
comparison ($C_{1,i}$) and control ($C_{2,i}$) stars in each AGN field
are given in Table~\ref{t_stars}. Coordinates are accurate to $\pm 3$
arcsec, while the accuracy of magnitudes varies between 0.05 and
0.10 mag, according to the photometric quality of each night. Hence,
these data are given just to allow the identification of these stars by future
observers, and should be used neither for astrometric purposes nor for
photometric calibration to the standard system.


\begin{table*} [tb]

\caption{Observational Results}
\begin{tabular}{lcccccrcc}

\noalign{\medskip}
\hline
\noalign{\smallskip}
~~Object  & UT Date & $q$ & $\sigma$ & $\Delta t$ & Variable? & $C$~~ &
$\Delta m_\mathrm{V}$ & $\langle V \rangle$  \\
       &      &      & mag   &   h      &           &     &
 mag  & mag   \\

\noalign{\smallskip}
\hline
\noalign{\smallskip}

0208$-$512 & 11/03/99 & 3 & 0.003 & 7.75 & V & 18.01 & 0.131 & $15.65 \pm 0.02$\\
           & 11/04/99 & 2 & 0.002 & 7.61 & V &  2.60 & 0.023 & $15.61 \pm 0.02$\\

0235$+$164 & 11/03/99 & 3 & 0.014 & 6.65 & V & 10.10 & 0.273 & $17.09 \pm 0.01$\\
           & 11/04/99 & 2 & 0.012 & 6.57 & V &  6.10 & 0.245 & $17.47 \pm 0.01$\\
           & 11/05/99 & 3 & 0.012 & 6.93 & V &  8.92 & 0.345 & $17.32 \pm 0.01$\\
           & 11/06/99 & 2 & 0.007 & 6.68 & V &  4.37 & 0.110 & $17.89 \pm 0.01$\\
           & 11/07/99 & 1 & 0.009 & 6.58 & V & 14.34 & 0.443 & $17.02 \pm 0.01$\\
           & 11/08/99 & 4 & 0.009 & 2.26 & V &  2.75 & 0.092 & $17.27 \pm 0.01$\\
           & 12/22/00 & 2 & 0.007 & 7.20 & V &  3.30 & 0.070 & $17.32 \pm 0.01$\\
           & 12/24/00 & 2 & 0.008 & 3.21 & V &  7.89 & 0.206 & $17.11 \pm 0.02$\\

0521$-$365 & 12/17/98 & 1 & 0.005 & 6.41 & V &  3.32 & 0.063 & $15.03 \pm 0.03$\\

0537$-$441 & 12/22/97 & 1 & 0.002 & 5.70 & V &  9.45 & 0.073 & $16.79 \pm 0.01$\\
           & 12/23/97 & 1 & 0.003 & 6.10 & V &  7.00 & 0.066 & $16.69 \pm 0.01$\\
           & 12/16/98 & 3 & 0.005 & 6.90 & V &  4.55 & 0.077 & $16.11 \pm 0.02$\\
           & 12/17/98 & 1 & 0.003 & 7.00 & V & 13.65 & 0.113 & $16.26 \pm 0.02$\\
           & 12/18/98 & 1 & 0.002 & 6.80 & V &  2.82 & 0.026 & $16.34 \pm 0.01$\\
           & 12/19/98 & 4 & 0.004 & 3.20 & V &  3.10 & 0.047 & $16.25 \pm 0.02$\\
           & 12/20/98 & 2 & 0.002 & 5.50 & NV & 2.25 & 0.025 & $16.04 \pm 0.02$\\
           & 12/21/98 & 2 & 0.002 & 6.50 & V  & 2.65 & 0.025 & $16.11 \pm 0.02$\\

           & 12/20/00 & 4 & 0.005 & 1.57 & NV & 0.51 & 0.008 & $14.30 \pm 0.05$\\
           & 12/21/00 & 1 & 0.002 & 7.04 & NV & 2.55 & 0.027 & $14.33 \pm 0.04$\\
           & 12/22/00 & 2 & 0.002 & 5.62 & NV & 1.39 & 0.012 & $14.33 \pm 0.03$\\
           & 12/23/00 & 1 & 0.002 & 4.99 & V  & 2.91 & 0.017 & $14.31 \pm 0.01$\\
           & 12/24/00 & 2 & 0.002 & 6.92 & V  & 3.60 & 0.031 & $14.22 \pm 0.01$\\

1226$+$023 & 04/08/00 & 1 & 0.005 & 7.60 & NV & 0.69 & 0.015 & $12.71 \pm 0.02$\\
           & 04/09/00 & 2 & 0.005 & 3.33 & NV & 1.16 & 0.030 & $12.67 \pm 0.02$\\

%
\label{t_results}
%
%

%
%

1229$-$021 & 04/11/00 & 1 & 0.008 & 8.10 & NV & 1.15 & 0.034 & $16.86 \pm 0.01$\\
           & 04/12/00 & 1 & 0.009 & 8.19 & NV & 1.19 & 0.046 & $16.86 \pm 0.01$\\

1243$-$072 & 04/08/00 & 1 & 0.020 & 7.52 & NV & 2.24 & 0.167 & $19.75 \pm 0.03$\\
           & 04/09/00 & 2 & 0.020 & 7.97 &  V & 3.00 & 0.187 & $19.74 \pm 0.03$\\

1253$-$055 & 06/08/99 & 4 & 0.024 & 3.81 & NV & 0.80 & 0.095 & $17.06 \pm 0.05$\\

1331$+$170 & 04/10/00 & 4 & 0.005 & 3.26 & NV & 1.17 & 0.026 & $16.34 \pm 0.04$\\

1334$-$127 & 04/11/00 & 1 & 0.009 & 8.82 & NV & 2.31 & 0.080 & $16.99 \pm 0.01$\\
           & 04/12/00 & 1 & 0.009 & 8.44 &  V & 2.67 & 0.081 & $17.16 \pm 0.01$\\

1424$-$418 & 06/04/99 & 4 & 0.023 & 5.99 & NV & 1.91 & 0.172 & $18.81 \pm 0.03$\\
           & 06/05/99 & 3 & 0.013 & 6.86 & NV & 1.86 & 0.081 & $18.85 \pm 0.02$\\

1510$-$089 & 04/29/98 & 3 & 0.004 & 3.74 & NV & 2.49 & 0.026 & $17.28 \pm 0.02$\\
           & 04/30/98 & 3 & 0.008 & 3.97 & NV & 1.45 & 0.042 & $17.26 \pm 0.02$\\
           & 06/06/99 & 4 & 0.009 & 7.23 & NV & 1.05 & 0.034 & $16.98 \pm 0.03$\\
           & 06/07/99 & 2 & 0.008 & 7.31 & NV & 1.27 & 0.042 & $16.97 \pm 0.06$\\

1606$+$106 & 07/23/01 & 1 & 0.005 & 4.59 & NV & 1.63 & 0.019 & $17.89 \pm 0.02$\\
           & 07/24/01 & 1 & 0.009 & 4.66 &  V & 3.08 & 0.092 & $17.81 \pm 0.02$\\

1622$-$297 & 06/04/99 & 4 & 0.021 & 6.75 & NV & 2.11 & 0.171 & $18.38 \pm 0.16$\\
           & 06/05/99 & 3 & 0.010 & 7.47 &  V & 3.21 & 0.132 & $18.34 \pm 0.16$\\

1741$-$038 & 06/06/99 & 4 & 0.025 & 7.92 & NV & 1.07 & 0.103 & $18.58 \pm 0.03$\\
           & 06/07/99 & 2 & 0.021 & 8.12 & NV & 1.73 & 0.151 & $18.57 \pm 0.07$\\

1933$-$400 & 07/23/01 & 1 & 0.008 & 9.05 & NV & 1.50 & 0.047 & $18.02 \pm 0.02$\\
           & 07/24/01 & 1 & 0.006 & 8.82 & NV & 2.28 & 0.071 & $18.00 \pm 0.02$\\

2022$-$077 & 07/25/01 & 1 & 0.010 & 6.94 &  V & 4.12 & 0.132 & $17.66 \pm 0.04$\\
           & 07/26/01 & 1 & 0.005 & 6.39 &  V & 4.89 & 0.091 & $17.06 \pm 0.04$\\

2155$-$304 & 07/27/97 & 3 & 0.002 & 7.00 & NV & 1.06 & 0.023 & $12.98 \pm 0.02$\\
           & 07/28/97 & 3 & 0.006 & 7.20 & NV & 0.73 & 0.027 & $12.98 \pm 0.02$\\

2230$+$114 & 07/23/01 & 1 & 0.004 & 5.74 & NV & 2.24 & 0.031 & $16.91 \pm 0.02$\\
           & 07/24/01 & 1 & 0.003 & 5.70 &  V &13.16 & 0.108 & $16.80 \pm 0.02$\\
           & 07/25/01 & 1 & 0.005 & 2.19 &  V & 6.80 & 0.093 & $16.39 \pm 0.01$\\

2320$-$035 & 07/25/01 & 1 & 0.005 & 6.18 & NV & 1.90 & 0.036 & $16.61 \pm 0.02$\\
           & 07/26/01 & 1 & 0.007 & 2.64 & NV & 1.38 & 0.026 & $16.58 \pm 0.02$\\

\noalign{\smallskip}
\hline

\end{tabular}

\end{table*}



The variability criterion adopted in the present work is the
99\%-confidence criterion used by \citet{JM97}, \citet{RCC99}, and many
others: a parameter $C=\sigma_\mathrm{T}/\sigma$ is introduced, where
$\sigma$ is the standard deviation of the control lightcurve and
$\sigma_\mathrm{T}$ the deviation of the target differential
lightcurve. A source can be then considered as variable at a 99\%
confidence if $C\geq 2.576$.


\section{Main results \label{s_mr}}

The results of our observations are shown in Table~\ref{t_results},
where we display, from left to right, the object name, the (Universal
Time) date of the observations, the quality parameter of each night
(see Sect.~\ref{s_obs}), the error determined from the standard
deviation of the comparison lightcurve, the duration of each observing
session, the classification of the source as variable (V) or
non-variable (NV) according to the scheme explained in the previous
section, the maximum magnitude fluctuation exhibited by the source in
the course of a single night, and, finally, the average $V$ magnitude
(in the standard system) for each night of observations. This last
parameter can be very different from what is listed in catalogs (see,
for instance, Table~\ref{t_obj}), since it changes with
time. Actually, it is clear that some sources that classify as NV at
microvariability timescales can be variable from night to night (i.e.,
at {\em inter\,}night timescales). Average magnitudes for nights with
quality parameter $q=3-4$ are probably affected by relatively high
systematic errors ($\lesssim 0.5$ mag) and hence should be taken with care.


\begin{table*}[tb]
\caption{Comparison and control stars.}

\begin{tabular}{lccclccc}

\noalign{\medskip}
\hline
\noalign{\smallskip}

 \multicolumn{4}{c}{Comparison stars} &
\multicolumn{4}{c}{Control stars} \\

Id. &  R.A. & Dec. & $V$ & Id. & R.A. & Dec. & $V$ \\

 & hs\,:\,min\,:\,s & $^\circ\,:\,'\,:\,''$ & mag & & hs\,:\,min\,:\,s & $^\circ\,:\,'\,:\,''$ & mag \\
\noalign{\smallskip}
\hline
\noalign{\smallskip}

\multicolumn{8}{c}{0208$-$512}\\

C$_{1,1}$ & 02:10:32.9 & -51:03:11 & 14.5 & C$_{2,1}$ & 02:10:36.2 & -50:59:31 & 14.6 \\
C$_{1,2}$ & 02:10:39.4 & -50:58:16 & 14.7 & C$_{2,2}$ & 02:10:57.0 & -51:01:52 & 15.7 \\
C$_{1,3}$ & 02:10:53.3 & -50:58:26 & 16.1 & C$_{2,3}$ & 02:10:43.2 & -51:03:25 & 16.2 \\

\noalign{\smallskip}
\multicolumn{8}{c}{0235$+$164}\\
C$_{1,1}$$^{a}$ & 02:38:38.4 & 16:38:19 & 16.8 & C$_{2,1}$ & 02:38:38.5 & 16:40:07 & 17.4 \\

\noalign{\smallskip}
\multicolumn{8}{c}{0521$-$365}\\
C$_{1,1}$ & 05:23:06.8 & -36:24:58 & 14.7 & C$_{2,1}$ & 05:22:40.4 & -36:26:13 & 15.2 \\
C$_{1,2}$ & 05:23:01.5 & -36:25:01 & 16.5 & C$_{2,2}$ & 05:22:49.8 & -36:29:24 & 16.0 \\
C$_{1,3}$ & 05:23:02.8 & -36:27:43 & 16.6 & C$_{2,3}$ & 05:22:53.6 & -36:23:24 & 16.6 \\

\noalign{\smallskip}
\multicolumn{8}{c}{0537$-$441}\\
C$_{1,1}$ & 05:38:55.7 & -44:06:22 & 14.8 & C$_{2,1}$ & 05:39:09.1 & -44:03:25 & 14.0 \\
C$_{1,2}$ & 05:39:06.7 & -44:03:11 & 15.2 & C$_{2,2}$ & 05:38:37.7 & -44:03:25 & 14.5 \\
C$_{1,3}$ & 05:38:29.9 & -44:04:59 & 15.2 & C$_{2,3}$ & 05:38:59.6 & -44:03:14 & 14.6 \\

\noalign{\smallskip}
\multicolumn{8}{c}{1226$+$023}\\
C$_{1,1}$$^{b}$ & 12:29:08.4 & 02:00:20 & 12.7 & C$_{2,1}$$^{c}$ & 12:29:03.1 & 02:03:19 & 13.5 \\
C$_{1,2}$ & 12:28:49.7 & 02:04:31 & 15.5 & C$_{2,2}$$^{d}$ & 12:29:02.9 & 02:02:17 & 14.9 \\

\noalign{\smallskip}
\multicolumn{8}{c}{1229$-$021}\\
C$_{1,1}$ & 12:32:03.6 & -02:21:57 & 17.2 & C$_{2,1}$ & 12:32:10.8 & -02:26:22 & 16.2 \\
C$_{1,2}$ & 12:32:03.1 & -02:26:05 & 17.3 & C$_{2,2}$ & 12:32:09.1 & -02:25:26 & 16.9 \\
C$_{1,3}$ & 12:32:13.7 & -02:24:54 & 17.8 & C$_{2,3}$ & 12:32:15.4 & -02:24:13 & 17.0 \\

\noalign{\smallskip}
\multicolumn{8}{c}{1243$-$072}\\
C$_{1,1}$ & 12:46:13.0 & -07:32:13 & 18.1 & C$_{2,1}$ & 12:46:01.2 & -07:32:04 & 18.6 \\
C$_{1,2}$ & 12:46:07.7 & -07:33:55 & 19.2 & C$_{2,2}$ & 12:46:06.7 & -07:28:40 & 19.0 \\
C$_{1,3}$ & 12:46:11.0 & -07:30:45 & 19.4 & C$_{2,3}$ & 12:46:01.2 & -07:29:16 & 19.3 \\

\noalign{\smallskip}
\multicolumn{8}{c}{1253$-$055}\\
C$_{1,1}$$^e$ & 12:56:26.6 & -05:45:21 & 16.4 & C$_{2,1}$ & 12:56:04.3 & -05:51:15 & 16.7 \\
C$_{1,2}$ & 12:56:04.1 & -05:49:11 & 17.8 & C$_{2,2}$ & 12:56:26.4 & -05:48:46 & 18.2 \\

\noalign{\smallskip}
\multicolumn{8}{c}{1331$+$170}\\
C$_{1,1}$ & 13:33:24.5 & 16:49:29 & 15.6 & C$_{2,1}$ & 13:33:33.4 & 16:47:47 & 16.0 \\
C$_{1,2}$ & 13:33:32.6 & 16:49:42 & 16.0 & C$_{2,2}$ & 13:33:12.7 & 16:50:43 & 16.3 \\

\noalign{\smallskip}
\multicolumn{8}{c}{1334$-$127}\\
C$_{1,1}$ & 13:37:34.6 & -12:58:16 & 17.9 & C$_{2,1}$ & 13:37:41.0 & -12:56:35 & 17.4 \\
C$_{1,2}$ & 13:37:48.0 & -12:54:22 & 18.0 & C$_{2,2}$ & 13:37:47.8 & -12:56:53 & 17.6 \\

\noalign{\smallskip}
\multicolumn{8}{c}{1424$-$418}\\
C$_{1,1}$ & 14:27:51.8 & -42:06:18 & 18.4 & C$_{2,1}$ & 14:27:58.6 & -42:08:06 & 17.6 \\
C$_{1,2}$ & 14:27:54.5 & -42:05:38 & 18.5 & C$_{2,2}$ & 14:27:48.0 & -42:08:42 & 18.4 \\
C$_{1,3}$ & 14:27:51.6 & -42:05:42 & 19.0 & C$_{2,3}$ & 14:27:54.5 & -42:09:14 & 18.6 \\

\noalign{\smallskip}
\multicolumn{8}{c}{1510$-$089}\\
C$_{1,1}$ & 15:12:53.0 & -09:05:41 & 16.7 & C$_{2,1}$ & 15:12:41.8 & -09:05:51 & 17.0 \\
C$_{1,2}$ & 15:13:00.0 & -09:05:42 & 17.5 & C$_{2,2}$ & 15:12:51.4 & -09:08:00 & 17.4 \\

\noalign{\smallskip}
\multicolumn{8}{c}{1606$+$106}\\
C$_{1,1}$ & 16:08:51.8 & 10:31:14 & 17.4 & C$_{2,1}$ & 16:08:48.0 & 10:27:27 & 17.4 \\
C$_{1,2}$ & 16:08:46.3 & 10:30:09 & 18.4 & C$_{2,2}$ & 16:08:51.8 & 10:32:40 & 17.6 \\
C$_{1,3}$ & 16:08:48.5 & 10:29:01 & 18.7 & C$_{2,3}$ & 16:08:48.5 & 10:31:57 & 17.7 \\

\noalign{\smallskip}
\hline
\noalign{\smallskip}
\end{tabular}

\label{t_stars}
\end{table*}


\setcounter{table}{2}
\begin{table*}
\caption{\textit{Continued.}}

\begin{tabular}{lccclccc}

\noalign{\medskip}
\hline
\noalign{\smallskip}

 \multicolumn{4}{c}{Comparison stars} &
\multicolumn{4}{c}{Control stars} \\

Id. &  R.A. & Dec. & $V$ & Id. & R.A. & Dec. & $V$ \\

 & hs\,:\,min\,:\,s & $^\circ\,:\,'\,:\,''$ & mag & & hs\,:\,min\,:\,s & $^\circ\,:\,'\,:\,''$ & mag \\
\noalign{\smallskip}
\hline
\noalign{\smallskip}

\noalign{\smallskip}
\multicolumn{8}{c}{1622$-$297}\\
C$_{1,1}$ & 16:28:46.2 & -29:51:53 & 18.2 & C$_{2,1}$ & 16:25:19.2 & -29:52:22 & 18.2 \\
C$_{1,2}$ & 16:24:56.9 & -29:51:02 & 18.3 & C$_{2,2}$ & 16:30:21.6 & -29:51:53 & 18.3 \\
C$_{1,3}$ & 16:25:53.4 & -29:52:43 & 18.4 & C$_{2,3}$ & 16:25:22.8 & -29:51:18 & 18.5 \\

\noalign{\smallskip}
\multicolumn{8}{c}{1741$-$038}\\
C$_{1,1}$ & 17:43:49.0 & -03:50:12 & 18.0 & C$_{2,1}$ & 17:44:01.4 & -03:49:22 & 18.0 \\
C$_{1,2}$ & 17:43:55.4 & -03:47:48 & 18.2 & C$_{2,2}$ & 17:44:01.9 & -03:50:23 & 18.5 \\

\noalign{\smallskip}
\multicolumn{8}{c}{1933$-$400}\\
C$_{1,1}$ & 19:37:10.3 & -39:55:44 & 17.7 & C$_{2,1}$ & 19:37:24.7 & -39:54:29 & 17.9 \\
C$_{1,2}$ & 19:37:20.6 & -39:59:02 & 18.0 & C$_{2,2}$ & 19:37:25.4 & -39:59:53 & 18.1 \\
C$_{1,3}$ & 19:37:20.9 & -39:57:40 & 18.2 & C$_{2,3}$ & 19:37:21.8 & -39:56:35 & 18.3 \\

\noalign{\smallskip}
\multicolumn{8}{c}{2022$-$077}\\
C$_{1,1}$ & 20:25:52.6 & -07:35:27 & 17.0 & C$_{2,1}$ & 20:25:44.2 & -07:37:57 & 17.2 \\
C$_{1,2}$ & 20:25:48.5 & -07:32:55 & 17.6 & C$_{2,2}$ & 20:25:49.4 & -07:35:01 & 17.6 \\
C$_{1,3}$ & 20:25:43.7 & -07:39:38 & 18.0 & C$_{2,3}$ & 20:25:45.1 & -07:37:20 & 18.0 \\

\noalign{\smallskip}
\multicolumn{8}{c}{2155$-$304}\\
C$_{1,1}$ & 21:58:43.7 & -30:17:17 & 14.1 & C$_{2,1}$ & 21:58:38.4 & -30:13:05 & 15.8 \\
C$_{1,2}$ & 21:58:35.8 & -30:13:34 & 14.9 & C$_{2,2}$ & 21:58:43.9 & -30:16:34 & 16.1 \\
C$_{1,3}$ & 21:58:57.8 & -30:13:26 & 15.4 & C$_{2,3}$ & 21:59:01.7 & -30:14:56 & 16.4 \\

\noalign{\smallskip}
\multicolumn{8}{c}{2230$+$114}\\
C$_{1,1}$ & 22:32:27.6 & 11:42:40 & 16.0 & C$_{2,1}$ & 22:32:32.9 & 11:42:45 & 16.3 \\
C$_{1,2}$ & 22:32:31.7 & 11:42:22 & 16.8 & C$_{2,2}$ & 22:32:45.8 & 11:42:01 & 17.1 \\
C$_{1,3}$ & 22:32:30.2 & 11:42:35 & 17.2 & C$_{2,3}$ & 22:32:48.5 & 11:44:44 & 17.2 \\

\noalign{\smallskip}
\multicolumn{8}{c}{2320$-$035}\\
C$_{1,1}$ & 23:23:39.6 & -03:18:58 & 15.9 & C$_{2,1}$ & 23:23:29.0 & -03:19:08 & 16.3 \\
C$_{1,2}$ & 23:23:25.9 & -03:19:34 & 17.0 & C$_{2,2}$ & 23:23:35.5 & -03:14:47 & 17.1 \\
C$_{1,3}$ & 23:23:33.8 & -03:15:33 & 17.3 & C$_{2,3}$ & 23:23:39.6 & -03:15:07 & 17.5 \\

\noalign{\smallskip}
\hline
\noalign{\medskip}

\noalign{$a$: Star 8 in \citealt{Smith85}}
\noalign{\medskip}
\noalign{$b$: Star E in \citealt{FTR98}}
\noalign{\medskip}
\noalign{$c$: Star G in \citealt{FTR98}}
\noalign{\medskip}
\noalign{$d$: Star B$'$ in \citealt{FTR98}}
\noalign{\medskip}
\noalign{$e$: Star 3 in \citealt{Rait98}}

\end{tabular}

\end{table*}

\begin{figure}
\includegraphics[width=\hsize]{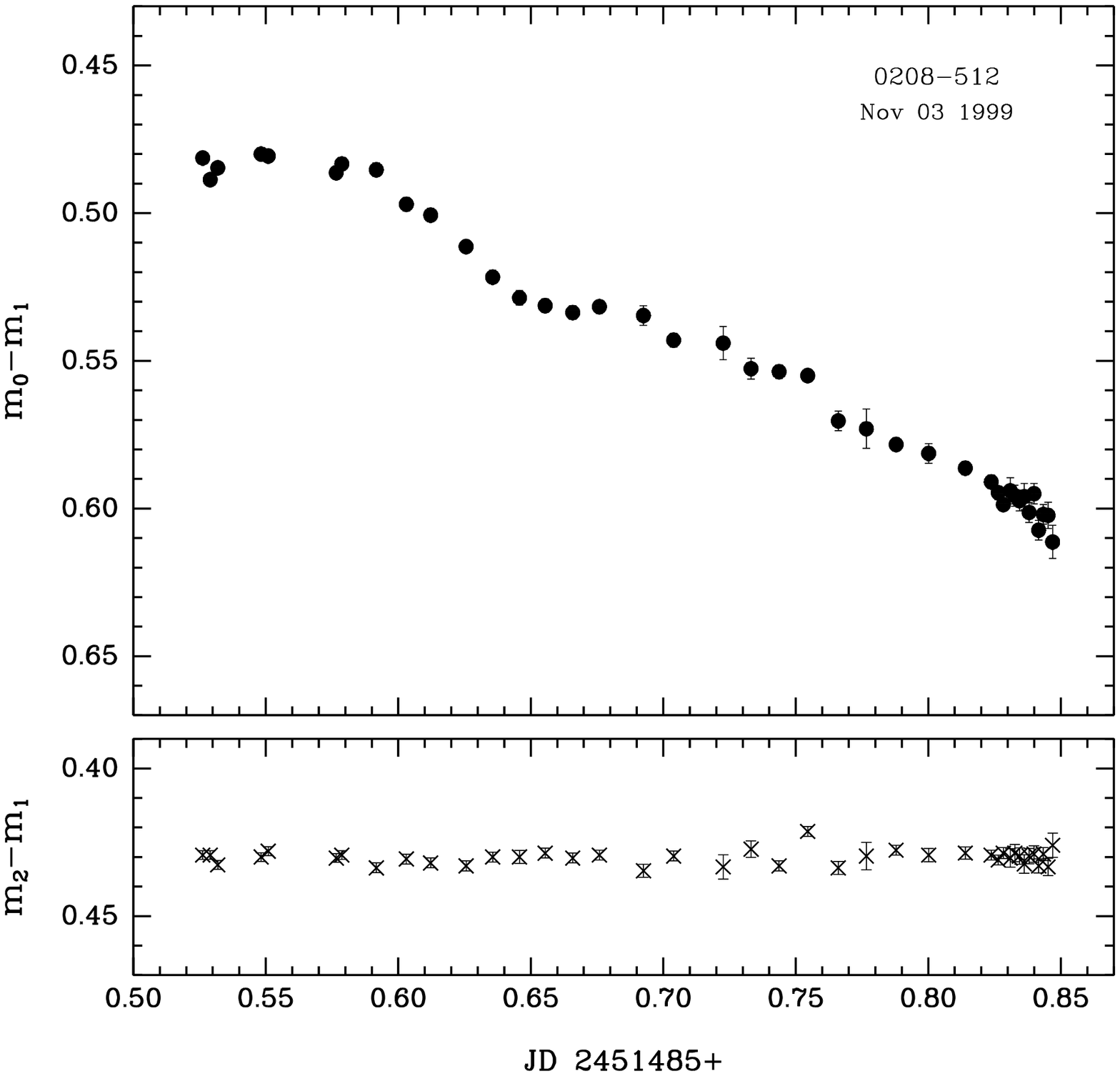}
\includegraphics[width=\hsize]{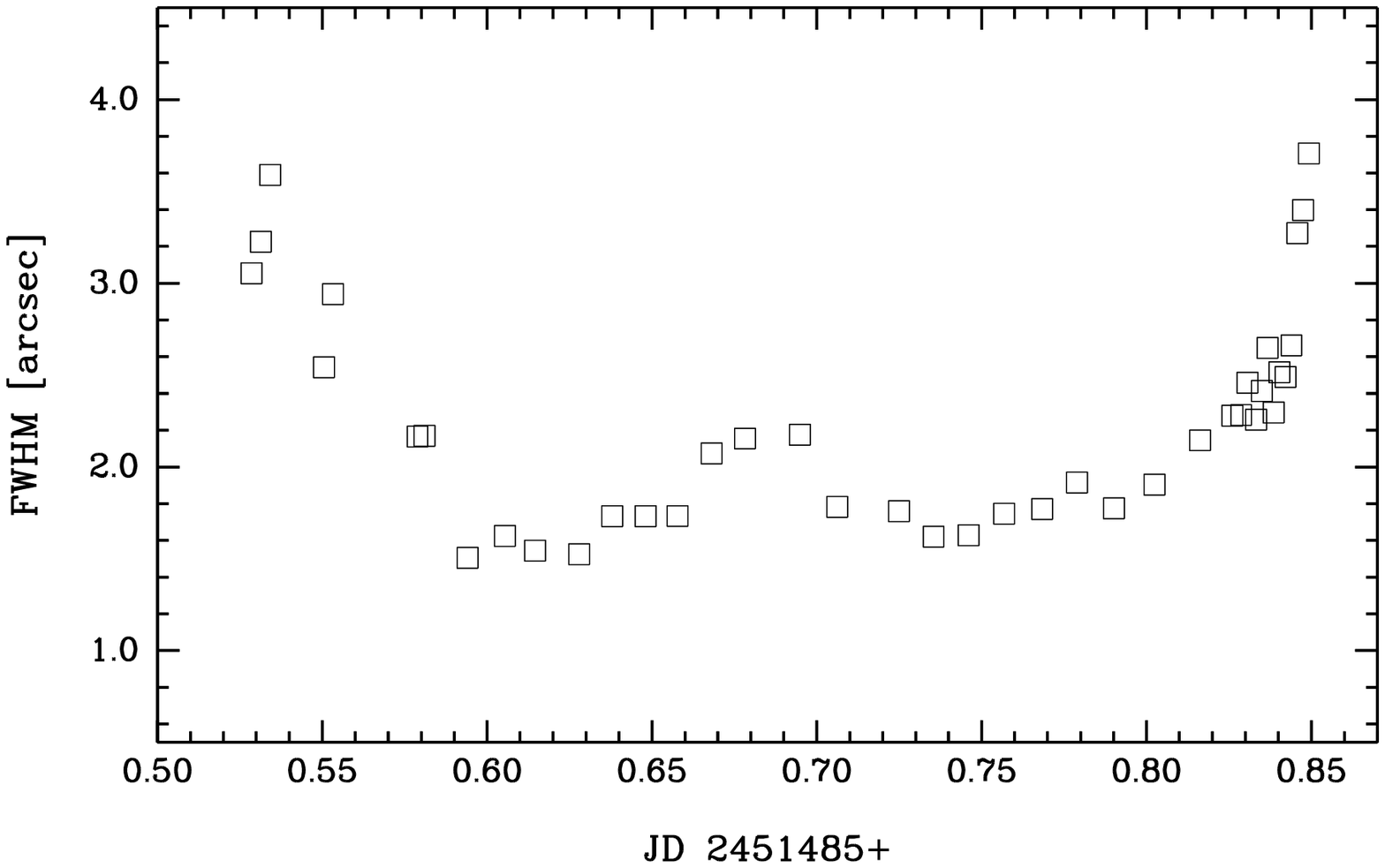}
 \caption{Differential lightcurve for 0208$-$512 on the night of
November 3rd, 1999, a typically variable blazar in the sample (filled
circles). Comparison stellar lightcurve is also shown (crosses).  In
the lower panel we show the atmospheric seeing evolution for this
night at CASLEO telescope (open squares). Notice the absence of
correlation.} \label{f_dlc0208}
\end{figure}

\begin{figure}
\includegraphics[width=\hsize]{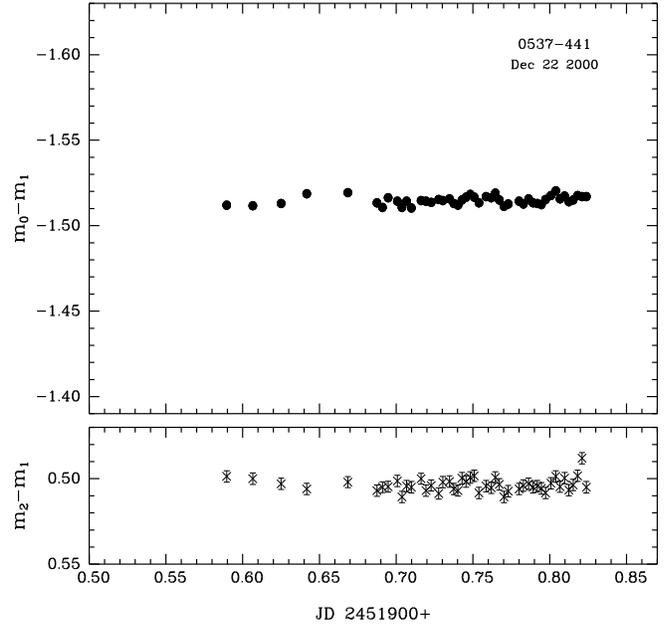}
\caption{Lightcurve for PKS\,0537$-$441 on the night of December
22nd, 2000. The source was not variable this particular
night. Comparison stellar lightcurve is also shown.}
\label{f_dlc0537}
\end{figure}

The most variable sources of our sample are the RBL objects
\object{AO\,0235$+$164} and \object{PKS\,0537$-$441}. We have
communicated separately on the outbursts observed on November 1999 and
December 1998, respectively, in these sources
\citep{RCC00a,RCC00b}. New data, from other observing sessions, are
added in this paper. Anyway, the duty cycle of AO\,0235$+$164 seems to
be close to 1. On the contrary, PKS\,0537$-$441 seems to switch
between states with high-level of variability and quiet states. The
intranight duty cycle for this source is $\sim 58.2$ \% (the {\em
inter}night value is about 81.6 \%). Many other gamma-ray blazars
were observed not to vary at all. For instance, the RLQ
\object{1510$-$089} was monitored on 4 epochs during
1998 and 1999 and was always found NV at intranight timescales, although it
was $\sim 0.3$ mag brighter in 1999. In Figures \ref{f_dlc0208} and
\ref{f_dlc0537} we show, just as examples, two lightcurves: one for a
variable source (\object{0208$-$512}, on the night of November 3rd,
1999, with $C=18.01$) and one for a non-variable source (the usually
considered ultra-variable RBL PKS\,0537$-$441, on the night of
December 22nd, 2000, with $C=1.39$), respectively. In Figure
\ref{f_dlc0208} (lower panel), we also show the evolution of the
atmospheric seeing during the observations. It can be clearly seen
that there is no correlation with the blazar differential
lightcurve. A seeing-variability correlation test was made for all
sources in the sample as recommended by \citet{CRC00} and implemented,
for example, by \citet{CC01}.

Complete lightcurves for all objects in the sample are published
electronically as Figures 2.1 to 2.20.

\begin{figure}
\includegraphics[width=\hsize]{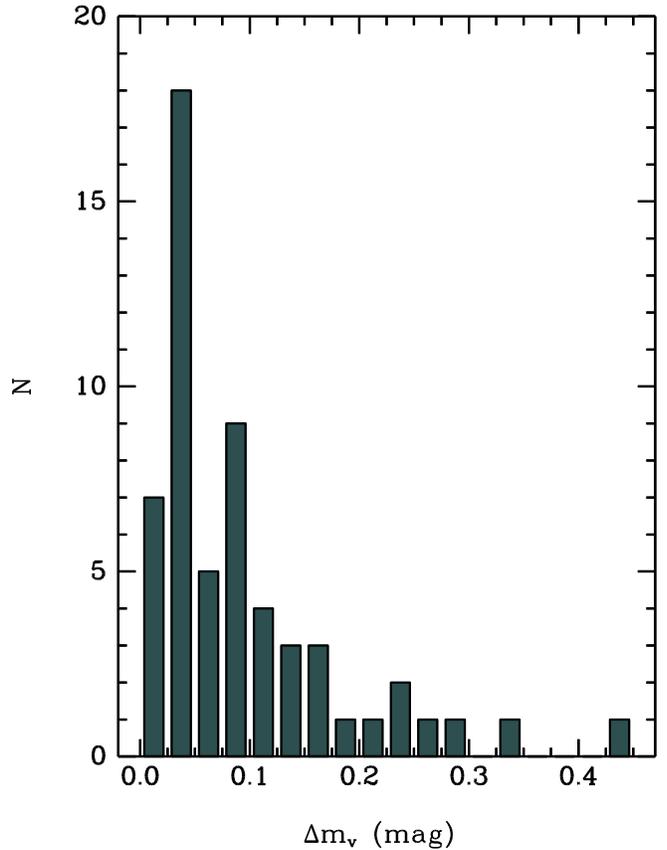}
\caption{\rm Histogram with
variability amplitudes for all objects in the sample.}
\label{h_dm}
\end{figure}

In Figure~\ref{h_dm} we present an histogram with the distribution of
variability amplitudes $\Delta m_V$ for all objects in the sample. The
highest microvariations in a single night are about 0.5 mag. Most of
the sources, however, present less violent changes.

\begin{figure}
\includegraphics[width=\hsize]{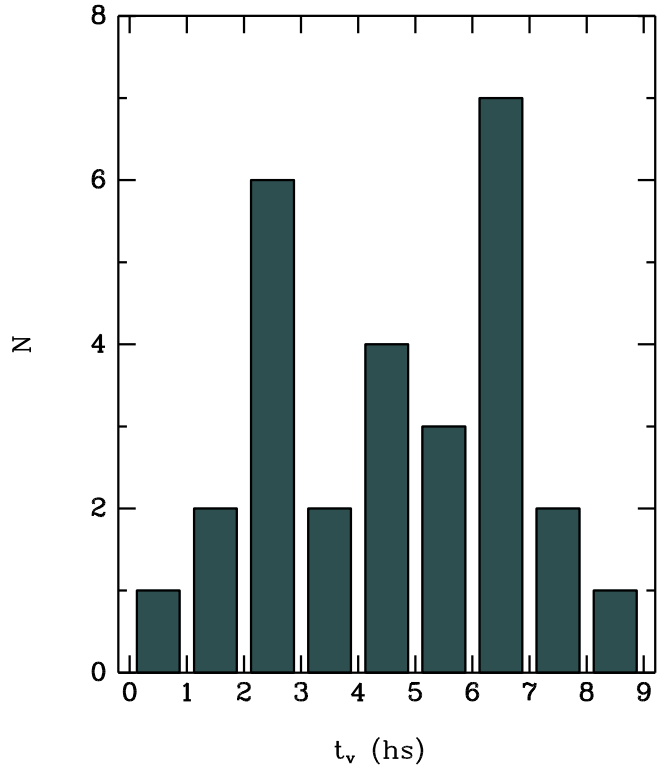}
\caption{Histogram with the microvariability timescales for
variable objects in the sample.} \label{h_tv}
\end{figure}

\begin{figure}
\includegraphics[width=\hsize]{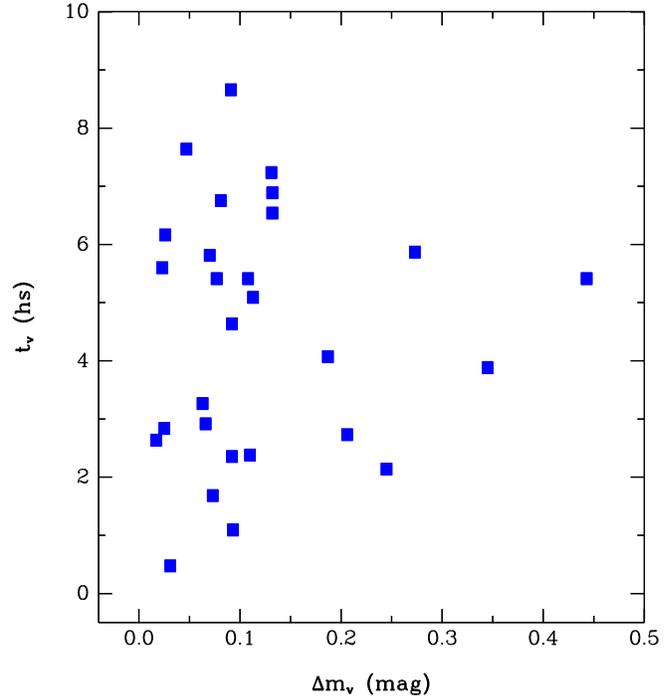}
\caption{Plot of the microvariability timescales vs.\ the
variability amplitude for objects in the sample that are variable
within a single night.} \label{t_dm}
\end{figure}

In Figure~\ref{h_tv} we show a similar histogram with the distribution
of the microvariability timescales. These are the timescales presented
by the largest amplitude variations occurred within a single night in
variable sources, and are defined as $t_{\rm v}=(1+z)^{-1}\Delta
F/(dF/dt)$, where $F$ is the flux density and the factor $(1+z)^{-1}$
is the cosmological correction. It can be seen that the shortest
timescales are of $\sim 1$ hour.Two peaks in the histogram indicate
that the preferred intranight timescales occur at 2-3 hours and 6-7
hours, although the second peak may be an artifact of the
observational sampling interval, since most sources were followed
during 6-7 hours per night. Figure~\ref{t_dm} shows a plot of $t_{\rm
v}$ vs.\ $\Delta m_V$, where it can be seen that the largest
intranight fluctuations occur with timescales in the range $2-6$
hours.

\section{Duty cycle of EGRET blazars \label{s_dc}}

\begin{figure}
\includegraphics[width=\hsize]{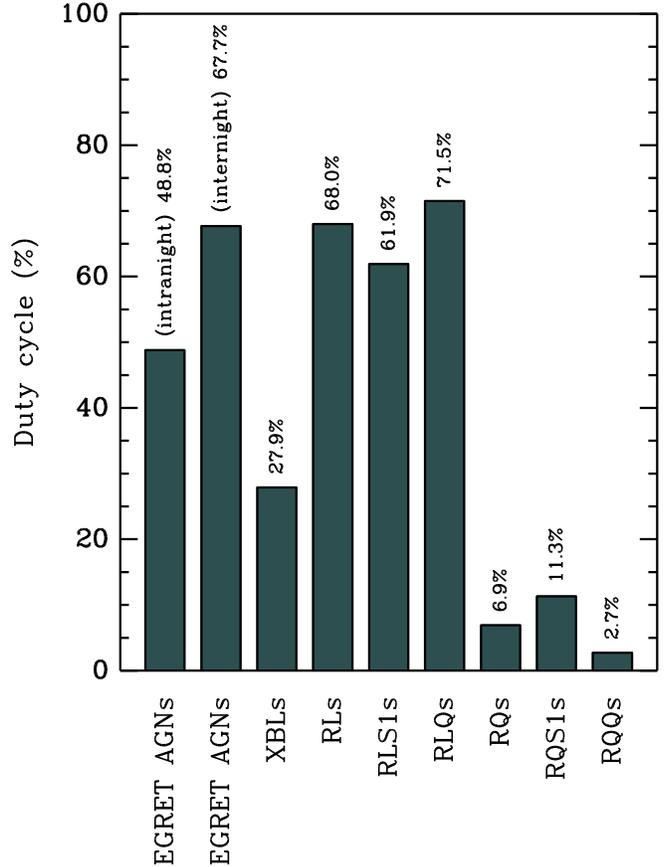}
\caption{Duty cycles for different types of AGNs, including
EGRET blazars, at strictly intranight timescales.  The duty cycle
for EGRET blazars at longer timescales (2 days) is also shown for
comparison. The group labeled as RQ includes both RQS1s and RQQs,
whereas RL objects are both RBLs and RLQs.} \label{h_dc}
\end{figure}

Duty cycles for objects of a given class can be roughly estimated as
\citep{RCC99}:
\begin{equation}
DC=100 \frac{\sum_{i=1}^n N_i (1/\Delta t_i)}{\sum_{i=1}^n (1/\Delta
t_i)}\%,
\label{DC}
\end{equation}
where $\Delta t_i=\Delta t_{i, \rm obs} (1+z)^{-1}$ is the
duration (corrected by redshift) of the $i$-st observing session
of a source of the class under study, and $N_i$ equals 0 or 1 if
the object was classified as NV or V during $\Delta t_i$,
respectively. Using this formula, \citet{RCC99} have estimated a
$DC$ of 71.5 \% for RBLs and RLQs, of 61.9 \% for radio-loud
Seyfert~1 galaxies (RS1s), of 27.9 \% for XBLs, and of only 2.7 \%
for radio-quiet QSOs. The class of the gamma-ray blazars detected
by EGRET includes objects classified as RBLs, RLQs and XBLs. If we
estimate, using Eq.~(\ref{DC}), the $DC$ for EGRET blazars as a
class, we get a value of 48.8 \%. This estimate is based on a
sample of 20 EGRET blazars and 57 independent observing sessions.
In Figure~\ref{h_dc} we present a graphic comparison between the $DC$ of
different classes of AGNs. We also indicate the value obtained for
the $DC$ of EGRET blazars when longer timescales (internight:
$\Delta t_i=2$ consecutive sessions) are considered. In this case
we get $DC = 67.7$ \%, and we are closer to the values presented
by the total class of flat-spectrum radio loud sources. The
conclusion seems to be that EGRET blazars are more active at
optical wavelengths on {\em inter}night than on {\em intra}night
timescales.

We emphasize that despite the time resolution of our observations is
very high, in any case we have detected the kind of events reported by
\citet{X99,X01,X02} for some EGRET blazars of our sample. These authors
claim detections of extreme events in objects included in our sample
like \object{1253$-$055} (3C279), 1510$-$089, and
AO\,0235$+$164 over timescales of a few minutes. For instance, they
report a variation of $\Delta V=1.17$ mag within 40 minutes on May 22,
1996 for 1253$-$055 and of $\Delta V=0.52$ mag within 10 minutes on
January 17, 1999 for AO\,0235$+$164. Our observations show no
indication of microvariability in objects like 1510$-$089 and, indeed,
strong variations in AO\,0235$+$164, an object with $DC\sim100$ \%, but
never over such extremely short timescales as reported by
\citeauthor{X99}
As communicated by \citet{RCC00a}, this latter source shows changes of
$\Delta V\sim0.5$ mag within a single night, but the well-resolved
lightcurves we have obtained present no indication of large-amplitude
changes on shorter (say less than 1 hour) timescales. It could be
argued that these particular sources could undergo extreme behaviour
from time to time and we have failed in its detection. But such a
behaviour was not observed in any of the 20 sources of our sample of
gamma-ray blazars. Since the minute-scale flares are claimed to be
present in many of the objects in the sample of \citeauthor{X99}, the
probability that we would have observed none of them along 57
observing sessions is extremely low. The mean $DC$ for minute-scale
microvariations seems to be $\sim 50$\,\% \citep{X01}, consequently the
probability of finding zero of such microvariations in our entire
campaign is $\sim 7\times 10^{-18}$.

A more likely alternative is that the discrepancies between both works
arise from different methods for error control.
\citet{X99,X01,X02} give no information on seeing fluctuations,
aperture size adopted for the photometric analysis, or light pollution
from the host.  Recently, \citet{CRC00} have demonstrated that strong
spurious variations in photometric observations of AGNs can occur even
in the absence of significant variability of the field stars, when the
AGN is embedded within a detectable host galaxy. These effects,
however, can be prevented through some simple techniques that we have
applied in our study (see Section 2 and \citealt{CRC00} for additional
details).

In a recent paper, \citet{D01} report to have also found
large-amplitude magnitude variations on very short timescales, even
after preventing against the effects of seeing induced light
contamination from the host galaxy.  However, a likely error source in
their data analysis, as well as in those of \citeauthor{X99}, lies in the fact
that many of the standard stars they used for calibration purposes are
significantly brighter than the corresponding AGN.

Remarkably, all the program objects reported to display fast, large
amplitude variations in \citeauthor{X99}'s papers and in \citet{D01},
are those for which the magnitude difference AGN $-$ standards is
largest (from $\sim 2.5$ and up to $\sim 5$ mag, standards always
brighter).  Our comparison and control stars, instead, typically
differ by only a few tenths of a magnitude from the corresponding AGN,
with a couple of extreme cases reaching a $\sim 1.5$ mag
difference. This ensures a correct matching of photometric and
aperture centering errors between AGN and comparison--control
stars. Let us mention that, because of their relatively high
brightnesses, all the standard stars in the field of AO\,0235$+$164
used by \citet{X01} are saturated in most of our images. The same is
true for 1510$-$089, for which \citet{D01} report a 1.72 mag fading in
27 minutes followed by a 2.00 mag brightening in 13 minutes, with a
variability parameter $C=59.6$. Note, however, that comparison star 1
in \citet{D01} is 5 magnitudes (i.e., 100 times in flux units)
brighter than the AGN, while star 2 is $\sim 3.3$ mag brighter than
the AGN [See \citet{Rait98} for coordinates and a finding chart for
these stars]. It is clear that, even with a high dynamic-range CCD, it is
not possible to properly expose the AGN in order to achieve a sufficiently
high signal-to-noise ratio without saturating the comparison star.

It is thus not surprising that \citet{D01} have found such a high variability
parameter $C$ for 1510$-$089. In fact, a simple error analysis
indicates that, with their observational setup, $C \simeq 17$ should
be expected from Poissonean noise alone, without considering possible
systematic errors from non-linearity or saturation, and even under
excellent atmospheric conditions and supposing a fairly dark sky
($\mu_R = 21\,\mathrm{mag\, arcsec}^{-2}$). For a site with a sky 1.5
mag\,arcsec$^{-2}$ brighter, a slightly larger value ($C \simeq 19$)
is obtained. Things get worse under non-photometric atmospheric
conditions, when the (faint) AGN will be relatively more affected than
the (bright) comparison star by the inevitable falloff in S/N ratio.

We conclude, contrary to previous claims, that the typical minimum
timescale for microvariations of EGRET blazars is of $\sim$ several
hours (not tens of minutes) and that the duty cycle for optical
microvariability in these objects peaks at timescales of $\sim 1-2$
days, in accordance with the short-term gamma-ray timescales observed
in several objects \citep[e.g.,][]{H96,H01,M01}. This result is
important since it implies that the optical and gamma-ray emitting
regions have similar sizes.

\section{Discussion \label{s_disc}}

Whereas the radio-to-UV continuum from blazars is usually interpreted
as synchrotron emission from energetic leptons in a relativistic jet,
the gamma-ray photons are thought to be the result of inverse Compton
scattering of soft seed photon fields by the same leptonic
population. The origin of these soft photons is not clear and
different possibilities have been discussed in the literature,
including external photon fields (accretion disk, broad line region)
and the own synchrotron photons produced in the jet (the so-called
synchrotron self-Compton model \hbox{--SSC--)}. The reader can find the
relevant references in the already cited paper by \citet{H01}.

Gamma-ray blazars are known to vary their flux on timescales as
short as $1-2$ days according to EGRET intensive monitoring of a
few selected sources like 3C279 and PKS 1406-076
\citep[e.g.,][]{W95,H01}. The shortest gamma-ray timescales are
about 8 hs, as reported for 3C279 by \citet{W98}. These timescales
are similar to the preferred timescales for optical
microvariability in EGRET blazars according to our study. This
seems to favor the idea that the optical and gamma-ray emission
are co-spatial, with sizes typically in range $10^{15}-10^{16}$
cm. Emission regions of such sizes are in accordance with the
constraints imposed by pair creation processes
\citep[e.g.,][]{BL95}. However, from non-simultaneous observations
the nature of the seed photons cannot be inferred. If simultaneous
optical/gamma observations clearly reveal the existence of
correlated microvariability with zero time lag, then SSC models
would be favored. On the contrary, non-simultaneous but well
correlated bursts could point out to external Compton models. In
particular, an optical burst preceding the gamma-ray flare (as in
\citealt{W95}) could be indicative of externally
rescattered/reprocessed synchrotron radiation of the jet
\citep{GM96}.

The fact that the minimum optical timescales for large microvariations
are of at least of several hours seems to agree with the suggestion
that in EGRET blazars the same population of relativistic particles is
responsible for both the optical and gamma-ray variable emission. If
the shortest optical timescales actually were of a few tens of
minutes, as claimed by Xie et al., then the optically emitting region
should be smaller than the Schwarzschild radius for objects like
3C279.

The most promising scenario to explain the production of both optical
and gamma-ray flares in blazars is perhaps the so-called shock-in-jet
model, where the increase in the flux density is the result of shocks
formed due to velocity irregularities in the relativistic flow of the
jet \citep[e.g.,][]{MG85,S01}. Particles in the shocked region cool
through synchrotron radiation at optical wavelengths and through
inverse Compton mechanism at gamma-ray energies.  Future simultaneous
optical/gamma-ray monitoring campaigns of blazars with high time
resolution using instruments like the forthcoming GLAST satellite will
help to solve the mystery of the origin of high-energy emission in
these objects and will provide elements to constrain the models of
nonthermal flares.

\begin{acknowledgements}
The authors acknowledge use of CCD and data acquisition system
supported under US National Science Foundation grant AST-90-15827 to
R.M. Rich. They are also very grateful to the CASLEO staff for their
kind assistance during the observations.  GER thanks the kind
hospitality of the Max-Planck-Institut f\"ur Kernphysik at Heidelberg,
where the final part of this work was done. This research was
supported by CONICET (PIP 0430/98), ANPCT (PICT 03-04881) and
Fundaci\'on Antorchas.
\end{acknowledgements}

\bibliographystyle{aa}

\end{document}